# Novel Metaknowledge-based Processing Technique for Multimedia Big Data clustering challenges

Nima Bari
Department of Computer Science at The George Washington University
Washington DC, USA
nbari@gwu.edu

Roman Vichr
Department of Data mining and Engineering
Exprentis, Inc.
Fairfax VA, USA
Roman.vichr@exprentis.com

Kamran Kowsari
Department of Computer Science at The George Washington University
Washington DC, USA
Kowsari@gwu.edu

Simon Y. Berkovich
Department of Computer Science at The George Washington University
Washington DC, USA
Berkov@gwu.edu

*Abstract*—Past research has challenged us with the task of showing relational patterns between text-based data and then clustering for predictive analysis using Golay Code technique. We focus on a novel approach to extract metaknowledge in multimedia datasets. Our collaboration has been an on-going task of studying the relational patterns between datapoints based on metafeatures extracted from metaknowledge in multimedia datasets. Those selected are significant to suit the mining technique we applied, Golay Code algorithm. In this research paper we summarize findings in optimization of metaknowledge representation for 23-bit representation of structured and unstructured multimedia data in order to be processed in 23-bit Golay Code for cluster recognition.

## I. INTRODUCTION

The latest focus of Big Data Analytics has been centered on content extraction and knowledge discovery [1], [4], [5]. We've done research developing the 23-bit metaknowledge template for Big Data clustering using mining and cluster techniques to confront the phenomena of Big Data. Our current research focuses on definition of datacube that is representative of data - each dimension of the cube - which can exceed 2 and even 3 dimensions - is an attribute of the data. This can theoretically allow us to converge specific datapoint as a result of an intended analysis. The 23-question Golay code template gives us a multi-dimensional platform for convergence towards predictive analysis.

## II. BIG MULTIMEDIA DATA CHALLENGES

Data retrieval from entities for the purpose of mining and analysis can prove difficult and is one of the challenging aspects of multimedia data clustering. Just as Osmar R. Zaiane and his collaborators explained utilizing the web to obtain data via crawling, such is the case then and now as a means for extracting the large amounts of data to create sub-repositories as a means of testing our hypotheses and theories.

[1] The authors of the Unified Framework for Representation, Analysis of Multimedia Content for Correlation and Prediction [3], Paul and Singh, highlighted some general challenges, which prove true when approaching any analysis of Big Data and especially Multimedia Big Data. First we must provide a structure where the multimedia data will be housed and represented effectively, i.e. prior to extracting any metaknowledge data. Next, given the amorphous nature of the data, we must consider both its multi-dimensional nature and conclude how to utilize the variety of attributes, have them coincide, and then apply methodologies [3].

## III. METAKNOWLEDGE PROCESSING FOR MULTIMEDIA BIG DATA

In our case selection of a concept benefitting the Golay algorithm, building the datacube is critical. Our task is rather simplified by using the proposed 23-questions metadata template. It allows us to determine significant data points according to 23 predetermined attributes and then cluster the set using Golay Coding [2]. Data is first subjected to our method of metafeature extraction in order to formulate an ontology. The ontology [6] allows us to represent inter-dependencies within one or more entities, therefore allowing for an unbroken collaborative approach. There are several challenges. The first challenge was to determine the specific classification algorithms applicable for our proposed ontology as noted in [7].

## IV. METHODOLOGY OF CONSTRUCTING AN INPUT

The methodology of extracting and deriving the datacube of metaknowledge representation is accomplished through statistical approach and methodology applied to structured data, while semantic methodology is applied to unstructured big data. In both cases, the layer of metaknowledge is constructed as the input Golay Code algorithm. This input is mapped into a representation of 23 bits (binary attribute vector) to create a hash value of index prior to processing by the code itself (Figure 1.). The Golay code processing includes implementation of fuzzy based on **H**amming **D**istance (**HD**) [17]. The Golay code guarantees linear processing time (**O**(n)) to build a hash table of indices and constant time (**O**(1)) for indices access (Figure1.). The attribute labeling No.1…No.23 is just a symbolic one and can represent Yes/No question or a presence of semantic element. In both cases it is then converted into binary representation of 0/1 (Figure 2.).

**GOLAY code utilizes 6 indices representing 86.5% of the index hash table**

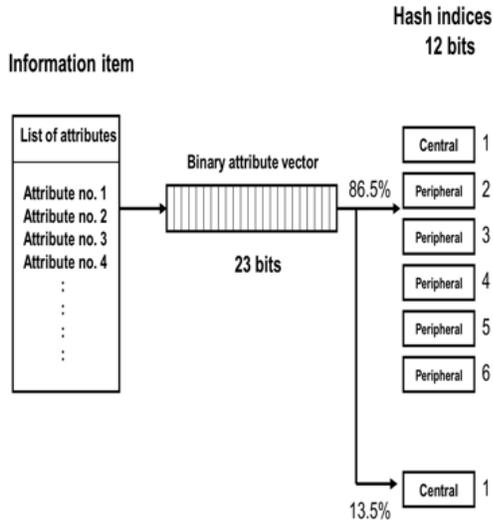

**Figure 1**: Golay Code 23 bit processing [11].

The result of the Golay code index lookup is represented by assigning a cluster label to the media record (the Golay code assigns only two label possibilities).

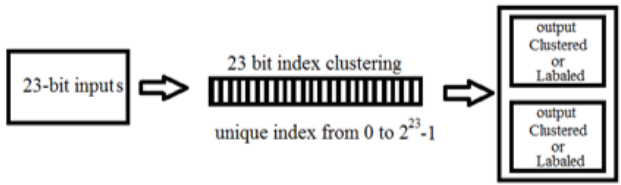

**Figure 2**: Golay Code 23bit index lookup and label resolution of clusters [11].

### V. (MULTIMEDIA) METAKNOWLEDGE REPRESENTATION AND STRUCTURED DATA

We applied our methodology to improve the extraction of metaknowledge representation from structured data, prior to applying the same to unstructured multimedia data. The classification algorithms we used are summarized in the table below, Table 1.

Decision trees were very effective in the validation of selecting metaknowledge attributes used to represent the bits of 23-bit record. DMRT was useful to derive threshold in metaknowledge single attribute representations.

| Num. | Metaknowledge and meta-feature discovery algorithms | |
|---|---|---|
| | *Algorithm* | *Benefits* |
| 1 | Scatter Plots (Correlation Coefficient) | > A quick assessment if the "average" pattern is linear, curved, or random<br>> If the trend is positive or negative association<br>> Strength of relationship<br>Identification of group of outliers (X,Y) |
| 2 | Statistical Relationship | > Statistical relationship in variation of possible values of X and X.<br>> Regression equation to describe the "best" line through the data and to predict Y based on X<br>> If linear relationship anticipated then describe the strength and direction |
| 3 | Decision Tree | > A fast learning curve, easy to test model and effective way to find "terminal" nodes.<br>> Purposeful test of quasi model by over-fitting to evaluate by adding/removing attributes.<br>> Assistance in deriving threshold driven questions to qualify answer Yes/No [1, 0]. |
| 4 | DMRT | > Discriminative classifier making validation of threshold setting in questions. |
| 5 | GLM | A quick determination of attribute (questions) impact and contribution to the binary outcome. |

**Table 1.**: Applied algorithms for metaknowledge feature discovery used in R [12].

The sequence of algorithms applied provides guidance to determine appropriateness and strength of metaknowledge [13], [14]. It also helps with assertions about values (thresholds of boundaries) of meta-feature template questions. These questions (as bits 1–23) are to be used as a foundation to represent the extracted metaknowledge in binary 23-bit word on the input of Golay Code processing.

The main advantage of this approach is to embed in processing a greedy loopback to minimize the error of given algorithms. For example, given GLM, then the AUC ( value minimization is the target of deriving subset of metaknowledge attributes that process can be embedded into greedy algorithm processing.

### VI. (MULTIMEDIA) METAKNOWLEDGE REPRESENTATION - SEMANTIC ONTOLOGY (UNSTRUCTURED DATA)

Applying the same methodology (described in V. Section) on unstructured data of multimedia file (in our collection html, .pdf, .jpg, .tiff) is much less effective as no structured data attributes with specific degrees of freedom are represented or easily extracted to construct datacubes. As Prof. Wen Gao from Peking University described in [17], text mining, which contains a lot of semantic information can contribute significantly on the retrieval process for content-based multimedia data such as video and audio. Therefore, naturally, we formulated a process to derive such metaknowledge utilizing semantic knowledge, so that multimedia can be categorized. In order to do that, a more generic characterization is done, i.e. not based on specific data value representations. It is based on the knowledge contained in the media file represented semantically. This turns out to be a more effective and accurate way to be useful in Golay Code processing when applied to multimedia file, and consequently applied to Big Data.

In order to test a solution to represent metaknowledge using semantic characterization of source files (hmtl, ms word, .pdf, .jpeg, etc.), we created a sample collection based on files obtained from [8], which is focused on the financial industry. It was easier to utilize the financial industry definition in terms of semantic structure and phrases.

First the semantic and generic definition was derived (see in Figure 3.) with the intent to define each semantic element as a [Yes/No] answer (per template for structured data) to be represented as binary for Golay Code processing. Therefore each file identified as having a semantic element match in its content or not in order to construct the datacube of metaknowledge. Such semantic element presence is then scored as 1 for present and 0 for not present and consequently processed with Golay Code.

The order of attributes within the datacube corresponds to the order of the most generic semantic elements to be placed first on the 23 bit record, followed by the most important ones to distinguish the records in clusters. In this case, we mean the first corresponds to the lowest bits of the 23bit input Golay Code record.

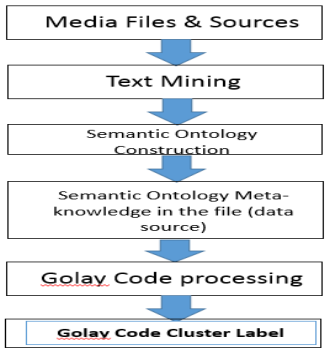

**Figure 3**: Derivation of Semantic metaknowledge per media files. The semantic metaknowldge is derived and extracted from media files and placed on the 23-bit record of Golay Code algorithm build using [9].

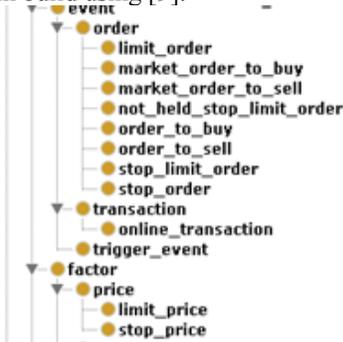

**Figure 4**: Semantic description of unstructured file (sample).

## VII. GOLAY CODE PROCESSING STRUCTURED DATA

The discovered metaknowledge was applied to the structured data of movie IMDB records [16] and those were consequently processed with Golay Code algorithm and assigned cluster label1 or label2. The best results in terms of TPR (**T**rue **P**ositive **R**ate) were obtained at HD=5, in Table 3. We then focused on optimization of metaknowledge attributes in Golay Code record to improve Spe**c**ificity (SPC) and **P**ositive **P**redictive **V**alue (PPV), which was accomplished. and summarized in Table 4. The **Acc**uracy (ACC) did not change dramatically in both cases.

| n | Run | Label1 | Label2 | HD | SRC_OSCARS WINS | Oscar Wins Label1 | Oscar Wins Label2 |
|---|---|---|---|---|---|---|---|
|   | 5 | 6413 | 3 | 7 | 251 | 251 | 0 |
| 6416 | 2 | 6241 | 175 | 6 | 251 | 171 | 76 |
|   | 5 | 5865 | 551 | 5 | 251 | 98 | 153 |
|   | 2 | 2705 | 3711 | 4 | 251 | 17 | 234 |
|   | 5 | 926 | 5490 | 3 | 251 | 0 | 251 |
|   | 2 | 22 | 6394 | 2 | 251 | 0 | 251 |

**Table 2**.: Golay Code processig of structured data characterized in metaknwoledge datacube.

Based on these result the confusion matrix [10] for structured data is represented in Table 3 (Hamming Distance = 5) and Table 4 ((Hamming Distance = 6).

| HD=5 | | | |
|---|---|---|---|
| TP | FP | TPR | 0.609562 |
| 153 | 398 | SPC | 0.935442 |
| 5767 | 98 | PPV | 0.277677 |
| TN | FN | ACC | 0.922693 |

**Table 3**.: Golay Code processing with HD=5.

| HD=6 | | | |
|---|---|---|---|
| TP | FP | TPR | 0.473469 |
| 116 | 46 | SPC | 0.992736 |
| 6287 | 129 | PPV | 0.716049 |
| TN | FN | ACC | 0.973396 |

**Table 4.:** Golay Code processing with HD=6.

As Table 4. shows, the focus to enhace metaknowledge attributes to increase SPC (**S**pecificity) lead to a drop in TPR (**T**rue **P**ositive **R**ate), because of the drop in True Positives (TP), but there was a significantly lower number of False Positives (FP). This was accomplished through the optimization of attributes into the metaknowledge datacube by placing the most significant metaknowledge attributes on the lowest bits of Golay code record input. At the same time we removed attributes duplicating the metaknowledge. The experiments show that the optimal HD (**H**amming **D**istance) to produce the best outcomes of clustering is with HD=4 and even better with HD=5.

## VIII. GOLAY CODE PROCESSING UNSTRUCTURED DATA

As we applied the process depicted in Figure 3, we produced semantic ontology metaknowledge representation of media unstructured records. The 23 bit record was constructed in such a way, that the lowest bits had generic semantic elements followed by the most distinct semantic element values. This is reflected in Table 5 (Run1). In this case the results were best for Hamming Distance (HD) = 2. An ideal result with True Positive Rate (TPR) = 1, i.e. all records were correctly clustered. Therefore for HD=2, Specificity (SPC) =1, Positive Predictive Value (PPV) =1 and Accuracy (ACC) also 1. All these terms refer to confusion matrix [15].

| RUN1 | |
|---|---|
| HD=2 | |
| TPR | 1 |
| SPC | 1 |
| PPV | 1 |
| ACC | 1 |

| HD=4 | | HD=6 | |
|---|---|---|---|
| TPR | 1 | TPR | 1 |
| SPC | 0.333333 | SPC | 0 |
| PPV | 0.8 | PPV | 0.727273 |
| ACC | 0.818182 | ACC | 0.727273 |

**Table 5.**: Golay Code processig (Expreriment RUN1)

Since the semantic metaknowledge is very concrete and specific, we obtained more accurate outcome of Golay Code clustering with smaller HD value (in this case HD=2, Run1 per Table 5.), rather than large HD valule.

In several subsequent runs, the order of the semantic metaknowlege attributes in 23-bit input representation was changed to understand better the impact on the final cluster assessment. Inaccurately assessed semantic metaknowledge has to be compensated for by increasing HD value. This case is demonstrated in results in Table 6. (Especially, the comparison of results per HD=2 and HD=4.). However, Specificity (SPC) drops to zero. An increase to HD=6 does not bring improvement in results.

| RUN3 | |
|---|---|
| HD=2 | |
| TPR | 0.875 |
| SPC | 0.75 |
| PPV | 0.875 |
| ACC | 0.833333 |

| HD=4 | | HD=6 | |
|---|---|---|---|
| TPR | 1 | TPR | 1 |
| SPC | 0 | SPC | 0 |
| PPV | 0.727273 | PPV | 0.727273 |
| ACC | 0.727273 | ACC | 0.727273 |

**Table 6.**: Golay Code processing (Experiment RUN3)

## IX. CONCLUSION AND FUTURE WORK

Our experiments were performed with extracted metaknowledge on structured and unstructured data to which we're applying Golay Code to categorize by label into cluster of records. Our approach had to be modified for media files to identify a proper metaknowledge, and therefore we introduced metaknowledge representation based on semantic ontology.

The experiments with unstructured data (media file) results show that an approach using ontology based meta-knowledge to process multimedia for clustering using Golay Code algorithm is a beneficial one. This is very useful in case of big data of media collections given the Golay code processing properties.

Our next research will also be focused into looking to utilize semantically driven metaknowledge attributes in order to dynamically derive processing rules, since the rules outcome can be defined as Yes/No (or True/False) outcome.

REFERENCES

[1] Zaïane, Osmar R., et al. "Mining multimedia data." Proceedings of the 1998 conference of the Centre for Advanced Studies on Collaborative research. IBM Press, 1998.

[2] Goldman-Segall, Ricki. "Challenges facing researchers using multimedia data: Tools for layering significance." ACM SIGGRAPH Computer Graphics 28.1 (1994): 48-51.

[3] Paul, S. Nissi, and Y. Jayanta Singh. "Unified framework for representation, analysis of multimedia content for correlation and prediction." Emerging Trends and Applications in Computer Science (ICETACS), 2013 1st International Conference on. IEEE, 2013.

[4] Huang, Tiejun, Yonghong Tian, Wen Gao, and Jian Lu. "Mediaprinting: Identifying multimedia content for digital rights management." (2007): 1-1..

[5] Chen, Chao, and Mei-Ling Shyu. "Clustering-based binary-class classification for imbalanced data sets." In Information Reuse and Integration (IRI), 2011 IEEE International Conference on, pp. 384-389. IEEE, 2011.

[6] Bari, Nima, Duoduo Liao, and Simon Berkovich. "Organization of Meta-knowledge in the Form of 23-bit Templates for Big Data Processing." In Computing for Geospatial Research and Application (COM. Geo), 2014 Fifth International Conference on, pp. 87-90. IEEE, 2014.

[7] Bari, Nima, Roman Vichr, Karmran Kowsari, Simon Berkovich. "23-Bit Metaknowledge Template towards Big Daata Knowledge Discovery and Management. The 2014 International Conference on Data Science and advanced Analytics. 2014.

[8] Financial Regulatory Agency – FINRA, www.finra.org

[9] Protégé, Stanford University, www.protege.org

[10] Confusion Matrix : http://en.wikipedia.org/wiki/Confusion_matrix

[11] Kamran Kowsari "Investigation of FuzzyFind Searching with Golay Code Transformations", M.Sc. Thesis, The George Washington University, Department of Computer Science, 2014

[12] R environment: cran.org

[13] Mathlab Group, MACSYMA reference manual, 1974, MIT.

[14] Confusion matrix: http://en.wikipedia.org/wiki/Confusion_matrix

[15] www.imdb.com

[16] Qixiang Ye , Wen Gao , Weiqiang Wang , Wei Zeng,, A robust Text Detection Algorithm, in Images and Video Frames, Fourth IEEE Pacific-Rim Conference On Multimedia, 2003

[17] Yammahi, Maryam and Kowsari, Kamran and Shen, Chen and Berkovich, and Simon. "An efficient technique for searching very large files with fuzzy criteria using the pigeonhole principle". IEEE, Computing for Geospatial Research and Application (COM. Geo), 2014 Fifth International Conference. 2014